# Thermodynamic Properties of Gaseous Plasmas in the Zero-Temperature Limit

Igor Iosilevskiy

*Joint Institute for High Temperature (Russian Academy of Science) Moscow, 125412, Russia*
*Moscow Institute of Physics and Technology (State University), Moscow, 141700, Russia*
ilios@orc.ru

Limiting structure of thermodynamic functions of gaseous plasmas is under consideration in the limit of extremely low temperature and density. Remarkable tendency, which was claimed previously, is carried to extreme. Both equations of state, thermal and caloric ones, obtain in this limit almost identical stepped structure ("ionization stairs") for plasma of single element, atomic or molecular. The point is that this limit ($T \to 0$; $n \to 0$) is carried out at fixed value of chemical potential for electrons or atoms. The same stepped structure is valid for plasma of mixtures or compounds. This structure appears within a fixed (negative) range of chemical potential of electrons (or atoms). It is bounded below by value of major ionization potential of given chemical element, and above by the value depending on sublimation energy of substance. Binding energies of all possible bound complexes (atomic, molecular, ionic and clustered) in its ground state are the only quantities that manifest itself in meaningful details of this limiting picture as location and value of every step. Energy of macroscopic binding (sublimation heat) supplements this collection and completes "intrinsic energy scale". All thermodynamic differential parameters (heat capacity, compressibility, *etc.*) obtain their remarkable δ-like structures in the zero-temperature limit ("thermodynamic spectrum"). All "lines" of these "spectrum" are centralized just at the elements of the intrinsic energy scale. The limiting EOS stepped structure of gaseous zero-Temperature isotherm is generic prototype of well-known "shell oscillations" in EOS of gaseous plasmas at low, but finite temperature. This limiting form of plasma thermodynamics could be used as a natural basis for rigorous deduction of quasi-chemical approach ("chemical picture") in frames of temperature (not density) asymptotic expansion around this reference system. The gaseous branch of zero-Temperature isotherm for energy depending on chemical potential could be naturally conjugated with associated branch of condensed state. The new gaseous branch of this generalized "cold curve" reflects in simple and schematic way all reactions (ionization, dissociation, phase transitions etc.) which are realized at the gas phase. At the same time new cold curve does not contain more meaningless portion of traditional cold curve representation, corresponding to thermodynamically unstable states. Remarkable limiting structure of thermodynamics of isotherm $T = 0$ for real substances manifests itself also in simplified classical models. Similar "ionization stairs", "thermodynamic spectrum" and modified "cold curve" are valid in modified one-component plasma on uniformly-compressible background, in two-component charged hard- and soft-spheres etc.

## 1. Ionization "stairs" in thermal and caloric EOS

Limiting structure of thermodynamic functions of gaseous plasmas is under consideration in the limit of extremely low temperature and density. Remarkable tendency, which was claimed previously [1-3], is carried to extreme. The point is that the discussed limit ($T \to 0$; $n \to 0$) is carried out at fixed value for chemical potential of electrons ($\mu_{el} = const$) or "atoms" ($\mu_a = Z\mu_{el} + \mu_i = const$) or "molecule" ($\mu_{m \leftrightarrow 2a} = 2\mu_a = const$) *etc*. In this limit both equations of state (EOS) thermal and caloric ones, obtain almost identical *stepped* structure ("ionization stairs" [3]) when one uses special forms for exposition of these EOS as a function of electron chemical potential: i.e. *PV/RT* for thermal EOS and $U - (3/2)PV$ for caloric EOS *vs.* $\mu_{el}$. Examples of this limiting structure are exposed at figures 1 and 2 for thermal and caloric EOS of lithium and helium plasmas [4-6]. For rigorous theoretical proof of existing the limit, which is under discussion (Saha-limit) in the case of hydrogen see [7, 8] and references therein.

The same *stepped* structure is valid in the zero-temperature limit for any molecular gases, for example for hydrogen (Fig. 3) [4][6]. It is also valid for simple plasma mixtures, for example for zero-temperature hydrogen-helium mixture with typical Sun-like composition (Fig.4). This limiting structure appears within a fixed (negative) range of $\mu_{el}$ ($\mu_{el}^{**} \geq \mu_{el} \geq \mu_{el}^{*}$). It is bounded below by value of major ionization potential of given chemical element ($\mu_{el}^{*} = -I_Z = -Z^2 Ry$) and above by the value depending on ionization potential and sublimation energy of substance $\{\mu_{el}{**} = -(\Delta_s H^o + I_1)/2\}$ [10]. Binding energies of all possible bound complexes (atomic, molecular, ionic and clustered) in its *ground state* are the *only* quantities that manifest itself in meaningful details of this limiting picture as location and value of every step. The energy of *macroscopic binding* – the heat of condensation at $T = 0$ – supplement this collection. At the same time there are *no* such *steps* for *exited states* of all bounded complexes (ions, atoms, molecules and clusters). Altogether, all energies mentioned above form *intrinsic energy scale* [3][10]. It could be considered as energy "passport" for any substance.

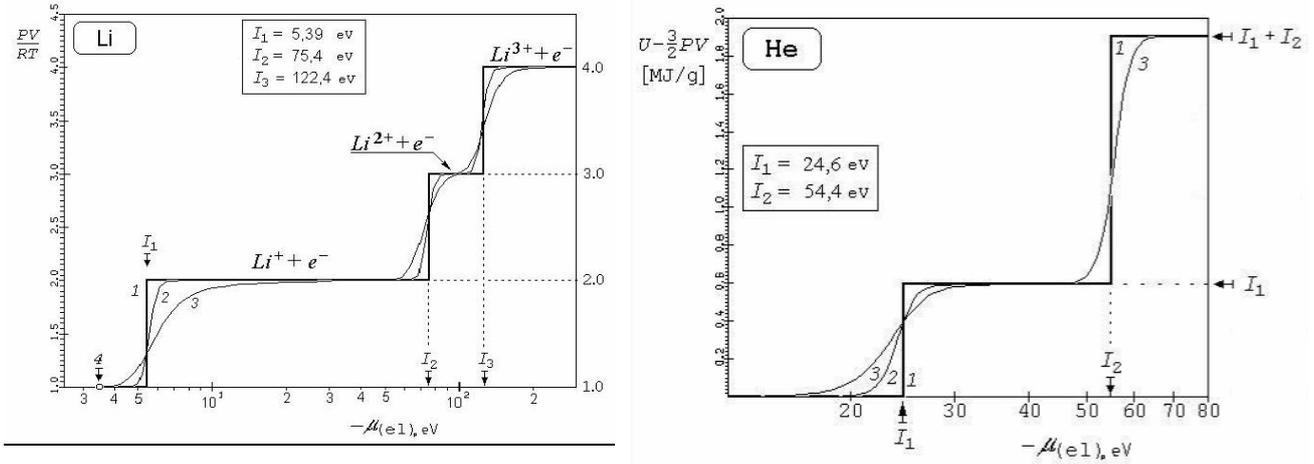

Figure 1. Thermal EOS of lithium plasma in quasi-chemical limit (figure from [4,6]). Compressibility factor $PV/RT \equiv P/(n_{Li}kT)$ as a function of (negative) value of electron chemical potential. *Notations*: *1* – isotherm $T = 0$; *2, 3* – isobars $P = 10^{-4}$ Pa and 1 MPa correspondingly; *arrows* – elements of lithium *intrinsic energy scale*: $I_1$, $I_2$, $I_3$ – lithium ionization potentials; *4* – point of condensation $\{(\mu_{el})^0 = -\{\Delta_s H^o/N) + I_1\}/2 = -3.51$ eV [3, 4]$\}$. (Isobars *2, 3* were calculated via code SAHA-IV [9] with neglecting of equilibrium radiation contribution)

Figure 2. Caloric EOS of helium plasma in quasi-chemical limit (figure from [6,10]. Complex $\{U - (3/2)PV\}$ as a function of (negative) value of electron chemical potential. *Notations*: *1,2,3* – isotherms $T = 0$, 10 000, 20 000 K; correspondingly; *arrows* – elements of helium *intrinsic energy scale*: $I_1$, $I_2$ – 1st and 2nd ionization potentials. (Isotherms *2,3* were calculated via code SAHA-IV [9] with neglecting of equilibrium radiation contribution)

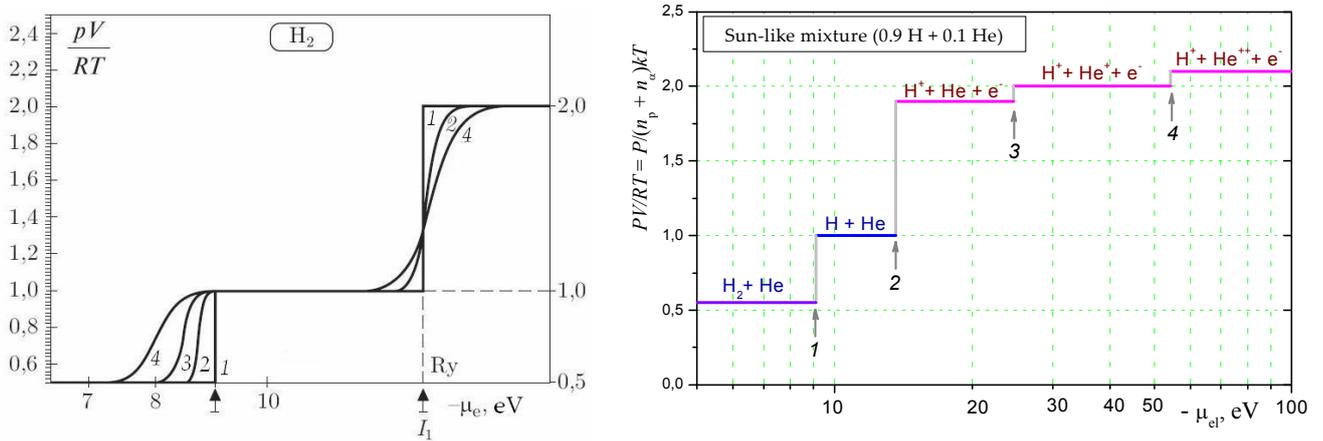

Figure 3. Thermal EOS of hydrogen plasma in quasi-chemical limit (figure from [4]) Compressibility factor $PV/RT \equiv P/(n_{H+} + n_H + 2n_{H2})kT$ as a function of (negative) value of electron chemical potential. *Notations*: *1,2,3,4* – isotherms $T = 0$, 1000, 2000, 4000 K correspondingly. $I_1 = $ Ry – hydrogen ionization potential. *5* – position of dissociation step at electron chemical potential scale $\{(\mu_{el})^D = -(D_2 + I)/2 = -9.04$ eV [3,4]$\}$. Isotherms *2-4* were calculated via code SAHA [11] with neglecting of equilibrium radiation contribution.

Figure 4. Thermal EOS of Sun-like hydrogen-helium mixture (0.9H + 0.1He) in quasi-chemical limit ($T \to 0$). Compressibility factor $PV/RT \equiv P/(n_{H.tot} + n_{He.tot})kT$ as a function of (negative) value of electronic chemical potential. *Notations*: *1,2* – positions of hydrogen dissociation and ionization steps, *3,4* – the same for the steps corresponding to the first and second ionization of helium (*1* – 9.04 eV, *2* – 13.6 eV, *3* – 24.6 eV, *4* – 54.4 eV).

## 2. Thermodynamic "spectrum"

In the zero-temperature limit all thermodynamic differential parameters (heat capacity, compressibility, *etc.*) obtain their remarkable δ-like structures ("thermodynamic spectrum" [3][10]). Both kinds of such "spectrum" became apparent: i.e. "emission-like spectrum" for heat capacity (Fig. 5) and "absorption-like spectrum" for the isentropic coefficient $\gamma_s \equiv (\partial \ln P / \partial \ln \rho)_S$ (Fig. 6). It should be stressed again that all "lines" of these "thermodynamic spectrum" are centralized just at the elements of the "*intrinsic energy scale*" – binding energies of ground states for all bound complexes in the system.

The limiting EOS stepped structure ("ionization stairs") of gaseous zero-Temperature isotherm is generic prototype of well-known "shell oscillations" in EOS of gaseous plasmas at low, but finite temperatures and non-idealities [2]. At the same time this limiting form of plasma thermodynamics could be used as a natural basis for rigorous deduction of well-known quasi-chemical approach ("chemical picture") in frames of asymptotic expansion around this reference system. The point is that this expansion must be provided on temperature at fixed chemical potential (see below), in contrast to the standard procedure of expansion on density at constant temperature [12, 1].

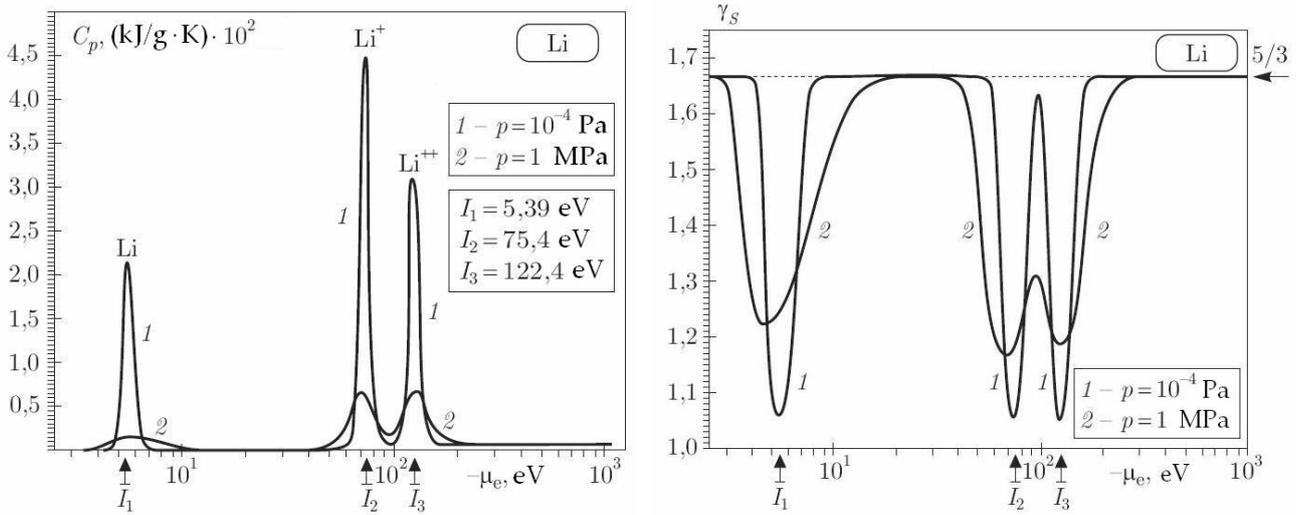

Figure 5,6. Limiting structure for differential thermodynamic quantities ("thermodynamic spectrum") in quasi-chemical limit $T \to 0$ (figure from [4, 6, 10]). Isobaric heat capacity (left panel) and isentropic coefficient $\gamma_s \equiv (\partial \ln P / \partial \ln \rho)_S$ (right panel) of lithium plasma as a function of (negative) value of electron chemical potential. *Notations*: *1, 2* – isobars $10^{-4}$ Pa and 1 MPa; *arrows* – elements of lithium intrinsic energy scale: $I_1, I_2, I_3$ – lithium ionization potentials. Ideal-gas value $(\partial \ln P / \partial \ln \rho)_S = 5/3$ is pointed out (Isobars *1,2* – SAHA code [9])

## 3. Jointed "cold curve" of matter (combined isotherm T = 0)

The gaseous branch of zero-temperature isotherm $U_0^{gas}(\mu)$ could be naturally conjugated with associated condensed branch $U_0^{crystal}(\mu)$. Due to the choice of chemical potential as a ruling parameter this combination creates complete and totally meaningful non-standard "cold curve" for any substance $\{U_0(\mu)$ instead of $U_0(\rho)\}$. The point is the appearance of stable thermodynamic gaseous branch for this "cold curve", which reflects schematically *all reactions* (ionization, dissociation etc.) and phase transitions which are realized at the system. Besides, the stable part of new combined "cold curve" it could be supplemented with additional *metatstable* branches, corresponding to overcooled vapor from gaseous part, and extended crystal from condensed part [13]. Another advantage of new representation for cold curve is natural identity of all transformations mentioned above (ionization, dissociation and phase transitions). It approves widely used interpretation of finite temperature ionization and dissociation as a "smoothed" phase transitions [10].

All present statements about remarkable limiting structure of thermodynamic functions in zero-temperature limit for single substances are valid also in application to chemical compounds. In this case mentioned above one-dimensional structures: "ionization stairs", modified "cold curve" and "thermodynamic spectrums" turn into more complicated two-dimensional constructions depending on not one, but two (or more) chemical potentials, $\mu_1$ and $\mu_2$, and composed also from discontinuity steps ("walls") and ideal-gas planes. Features and properties of such limiting structures are non-investigated at the moment. Study of such structures is in progress (to be published)

## 4. Zero-temperature thermodynamics in simplified Coulomb models

Remarkable limiting structure of thermodynamics for real substances, which is under discussion, manifests itself also in simplified classical models. Similar "ionization stairs", "thermodynamic spectrum" and modified "cold curve" was predicted [3,13,10] for modified one-component ionic model on uniformly-compressible compensating background, OCP(~) [14,15] and for two-component classical ionic model with Glauberman's [16] potential $\{V_{ij}(r) \equiv Z_i Z_j e^2[1 - \exp(-r/\sigma)]/r\}$ [4] and for classical charged hard- and soft-spheres models. In the first model OCP(~) there is no electron-ionic associations on definition. The only transformations permitted in the model are three 1st-order phase transitions between solid, liquid and gas-like states. It leads to non-standard cold curve in OCP(~) with sublimation jump and metatstable regions [4, 17].

## 5. The problem of correct derivation for "chemical picture"

New representation for cold curve (isotherm $T=0$), which is introduced in present paper, has advantage for solution of theoretical problem of correct deducing of quasi-chemical representation ("chemical picture") – ensemble of "free" elementary and complex particles: ions, electrons, atoms and molecules, with relatively weak *effective* interaction, from rigorous physical representation (ensemble of nuclei and electrons with strong Coulomb interaction). The well-known and widely accepted traditional theoretical approach uses asymptotic expansion in terms of *activities* (or partial densities) at *constant temperature* (for example [12]). The basic statement of present work [3,5,10] is that in contrast to this traditional approach the new supplementary one should be developed as more adequate for thermodynamics of low-temperature multi-component plasmas with ample composition of bounded complexes of arbitrary complexity. Main point of this approach should be systematic asymptotic expansion of all thermodynamic functions by functions of *temperature* in the limit $T \to 0$ at *fixed value* of *chemical potential*(*s*).

$$Y(\mu, T) = Y_0(\mu) + Y_1(\mu, T) + Y_2(\mu, T) + ... + Y_k(\mu, T) + ... + Y_{k+1}(\mu, T) + ... \quad (1)$$
$$\{Y_{k+1}(\mu, T)/Y_k(\mu, T)\} \to 0 \quad (T \to 0; \mu = const)$$

It should be stressed that such desirable approach should develop asymptotic expansion in the limit $T \to 0$ simultaneously for *all values* of chemical potential within mentioned above "energy scale" from the full ionization state of nuclei and electrons ($\mu_{el} \sim -I_Z$) up to the saturation point in atomic and molecular mixture and including the intermediate regions for *all stages* of ionization and dissociation. Both discussed above ionization "stairs" in thermal and caloric EOS are natural temperature independent zero-order terms $Y_0(\mu)$ in such global asymptotic expansion (1). Such asymptotic expansion by the functions of temperature $\lambda_i(\mu) \sim \exp\{-const_i(\mu)/T\}$ was rigorously developed for thermal EOS of hydrogen $P(T,\rho)$ in the case of partially ionized electron-ion-atomic hydrogen (SAHA-limit, see [7, 8] and references therein). The main claim of present paper is existence of such asymptotic expansion for *any* substance and in *whole range* of chemical potential within the energy scale of the substance, including creation of *any bound complexes*, multi-electronic ions, atoms, molecules etc. up to condensation point.

### Acknowledgments

The work was supported by Grants ISTC 3755, CRDF № MO-011-0, by Scientific Program of Russian Academy of Science "Physics of Extreme States of Matter" and by MIPT Education Center "Physics of High Energy Density Matter".